# Large-scale compression of genomic sequence databases with the Burrows-Wheeler transform

Anthony J. Cox,[1,][*] Markus J. Bauer,[1] Tobias Jakobi[2] and Giovanna Rosone[3]

[1]Computational Biology Group, Illumina Cambridge Ltd., Chesterford Research Park, Little Chesterford, Essex CB10 1XL, United Kingdom
[2]Computational Genomics, CeBiTec, Bielefeld University, Bielefeld, Germany
[3]University of Palermo, Dipartimento di Matematica e Informatica,Via Archirafi 34, 90123 Palermo, Italy

## ABSTRACT

**Motivation:**

The Burrows-Wheeler transform (BWT) is the foundation of many algorithms for compression and indexing of text data, but the cost of computing the BWT of very large string collections has prevented these techniques from being widely applied to the large sets of sequences often encountered as the outcome of DNA sequencing experiments. In previous work (Bauer *et al.* (2011)), we presented a novel algorithm that allows the BWT of human genome scale data to be computed on very moderate hardware, thus enabling us to investigate the BWT as a tool for the compression of such datasets.

**Results:**

We first used simulated reads to explore the relationship between the level of compression and the error rate, the length of the reads and the level of sampling of the underlying genome and compare choices of second-stage compression algorithm.

We demonstrate that compression may be greatly improved by a particular reordering of the sequences in the collection and give a novel 'implicit sorting' strategy that enables these benefits to be realised without the overhead of sorting the reads. With these techniques, a 45× coverage of real human genome sequence data compresses losslessly to under 0.5 bits per base, allowing the 135.3Gbp of sequence to fit into only 8.2Gbytes of space (trimming a small proportion of low-quality bases from the reads improves the compression still further).

This is more than 4 times smaller than the size achieved by a standard BWT-based compressor (`bzip2`) on the untrimmed reads, but an important further advantage of our approach is that it facilitates the building of compressed full text indexes such as the FM-index (Ferragina and Manzini (2000)) on large-scale DNA sequence collections.

**Availability:**

Code to construct the BWT and SAP-array on large genomic data sets is part of the `BEETL` library, available as a github respository at `git@github.com:BEETL/BEETL.git`.

**Contact:** acox@illumina.com

## 1 INTRODUCTION

In this paper we present strategies for the lossless compression of the large number of short DNA sequences that comprise the raw data of a typical sequencing experiment.

Much of the early work on the compression of DNA sequences was motivated by the notion that the compressibility of a DNA sequence could serve as a measure of its information content and hence as a tool for sequence analysis. This concept was applied to topics such as feature detection in genomes (Grumbach and Tahi (1994); Rivals *et al.* (1996); Milosavljevic and Jurka (1993)) and alignment-free methods of sequence comparison (Chen *et al.* (2002)) - a comprehensive review of the field up to 2009 is given by Giancarlo *et al.* (2009) . However, Grumbach and Tahi in 1994 have been echoed by many subsequent authors in citing the exponential growth in the size of nucleotide sequence databases as a reason to be interested in compression for its own sake. The recent and rapid evolution of DNA sequencing technology has given the topic more practical relevance than ever.

The outcome of a sequencing experiment typically comprises a large number of short sequences - often called 'reads' - plus metadata associated with each read and a 'quality score' that estimates the confidence of each base. Tembe *et al.* (2010) and Deorowicz and Grabowski (2011) both describe methods for compressing the FASTQ file format in which such data is often stored. The metadata is usually highly redundant, whereas the quality scores can be hard to compress, and these two factors combine to make it hard to estimate the degree of compression achieved for the sequences themselves. However, both schemes employ judicious combinations of standard text compression methods such as Huffman and Lempel-Ziv, with which it is hard to improve substantially upon the naive method of using a different 2-bit code for each of the 4 nucleotide bases. For example, GenCompress (Chen *et al.* (2002)) obtains 1.92 bits per base (henceforth bpb) compression on the *E.coli* genome.

An experimenter wishing to sequence a diploid genome such as a human might aim for 20-fold average coverage or more, with the intention of ensuring a high probability of capturing both alleles of any heterozygous variation. This oversampling creates an opportunity for compression that is additional to any redundancy inherent in the sample being sequenced. However, in a whole-genome shotgun experiment, the multiple copies of each locus are randomly dispersed among the many millions of reads in the dataset, making this redundancy inaccessible to any compression method that relies on comparison with a small buffer of recently-seen data.

[*]to whom correspondence should be addressed

This can be addressed by 'reference-based' compression (Kozanitis *et al.* (2010); Fritz *et al.* (2011)), which saves space by sorting aligned reads by the position they align to on a reference sequence and expressing their sequences as compact encodings of the differences between the reads and the reference. However this is fundamentally a lossy strategy that achieves best compression by retaining only reads that closely match the reference, limiting the scope for future reanalyses such as realignment to a refined reference (containing, perhaps, ethnicity specific haplotypes (Dewey *et al.* (2011)) or any sort of *de novo* discovery on reads that did not align well initially. Moreover, as Yanovsky (2011) points out, a reference-based approach is inapplicable to experiments for which a reference sequence is not clearly defined (metagenomics) or entirely absent (*de novo*).

Yanovsky (2011) describes a lossless compression method `ReCoil` for sets of reads that works in external memory (*i.e.* via sequential access to files held on disk) and is therefore not constrained in scale by available RAM. A graph of similarities between the reads is first constructed and then each read is expressed as a traversal on that graph, encodings of these traversals then being passed to a general-purpose compressor (`ReCoil` uses `7-Zip`[1]).

The two-stage nature of this procedure is shared by the family of compression algorithms based on the Burrows-Wheeler transform (BWT). The BWT is simply a permutation of the letters of the text and so is not a compression method *per se*. Its usefulness for compression stems from the facts that the BWT tends to be more compressible than its originating text (since it tends to group symbols into 'runs' of like letters, which are easy to compress) and, remarkably, that the originating text can be reconstructed from it (thus allowing the BWT to represent the originating text without loss of information). Once generated, the BWT is compressed by standard techniques: a typical scheme would follow an initial move-to-front encoding with run length encoding and then Huffman encoding.

The widely-used BWT-based compressor `bzip2`[2] divides a text into blocks of (at most, and by default) 900 kbytes and compresses each separately, so is only able to take advantage of local similarities in the data. Mantaci *et al.* (2005) gave the first extension of the BWT to a collection of sequences and used it as a preprocessing step for the simultaneous compression of the sequences of the collection. They show that this method is more effective than the technique used by a classic BWT-based compressor, because one could potentially access redundancy arising from these long-range correlations in the data. Until recently, computing the BWT of a large collection of sequences was prevented from being feasible on very large scales by the need either to store the suffix array of the set of reads in RAM (requiring 400Gbytes of RAM for a dataset of 1 billion 100-mer reads) or to resort to "divide-and-conquer then merge" strategies at considerable cost in CPU time. However, in Bauer *et al.* (2011), three of the present authors described fast and RAM-efficient methods capable of computing the BWT of sequence collections of the size encountered in human whole genome sequencing experiments, computing the BWT of collections as large as 1 billion 100-mers.

Unlike the transformation in Mantaci *et al.* (2005), the algorithms in Bauer *et al.* (2011) require ordered and distinct 'end-marker' characters to be appended to the sequences in the collection, making the collection of sequences an ordered multiset, *i.e.* the order of the sequences in the collection is determined by the lexicographical order of these end-markers. It is easy to see that the ordering of sequences in the collection can affect the compression, since the same or similar sequences might be distant in the collection. We will outline different ordering strategies for the input sequences and will show their effect on simulated and real data.

We have created an open-source C++ library `BEETL` that makes it practical to compute the BWT of large collections of DNA sequences and thus enables the redundancy present in large-scale genomic sequence datasets to be fully exploited by generic second-stage compressors such as `bzip2` and `7-Zip`. We use simulated read collections of up to $60\times$ redundancy to investigate the effect of genome coverage, sequencing error and read length on the level of compression. Furthermore, we show the effect of read trimming and choices of second-stage compressors on the level of compression. Finally, we describe an extension of our method that implicitly reorders sequences in the collection so as to drastically improve compression.

## 2 METHODS

Consider a string $s$ comprising $k$ symbols from some finite ordered alphabet $\Sigma = \{c_1, c_2, \ldots, c_\sigma\}$ of size $\sigma$. We mark the end of $s$ by appending an additional symbol $ that will be considered to be lexicographically smaller than any symbol in $\Sigma$. We can build $k + 1$ distinct *suffixes* from $s$ by starting at different symbols of the string and continuing rightwards until we reach $. If we imagine placing these suffixes in alphabetical order, then the *Burrows-Wheeler transform* of $s$ (Burrows and Wheeler (1994)) can be defined such that the $i$-th element of the BWT is the symbol in $s$ that precedes the first symbol of the $i$-th member of this ordered list of suffixes. Each symbol in the BWT therefore has an *associated suffix* in the string. We recommend Adjeroh *et al.* (2008) for further reading.

One way to generalize the notion of the BWT to a collection of $m$ strings $S = \{s_1, \ldots, s_m\}$ (see also Mantaci *et al.* (2005)) is to imagine that each member $s_i$ of the collection is terminated by a distinct end marker $_i$ such that $_1 < \cdots < $_m$. Moreover, we assume that the end markers are lexicographically smaller than any symbol in $\Sigma$. In Bauer *et al.* (2011), we give two related methods for computing the BWT of large collections of DNA sequences by making use of sequential reading and writing of files from disk. The first variant `BCR` requires 14 bytes of RAM for each sequence in the collection to be processed (and is hence capable of processing over a billion reads in 16Gbytes of RAM), whereas the second variant `BCRext` uses negligible RAM at the expense of a larger amount of disk I/O.

To understand how a BWT string might be compressed, we can think of it as the concatenation of a set of 'runs' of like letters, each of which can be described by its constituent symbol $c$ plus an integer $i$ denoting the number of times $c$ is repeated. We assume

---

all runs are maximal, *i.e.* they do not abut another run of the same character. Intuitively, for two strings of the same length, we expect the string that consists of fewer (and hence, on average, longer) runs to compress to a smaller size.

Given $S$, our strategy is to search for permutations $S \rightarrow S'$ of the sequences in the collection such that the BWT of the permuted collection $S'$ can be more readily compressed than the BWT of $S$. For the BWT of $S$, we define a bit array called the *SAP-array* (for 'same-as-previous') whose elements are set to $\mathbf{1}$ if and only if the suffixes associated with their corresponding characters in the BWT are identical (their end markers excluded) to those associated with the characters that precede them. Thus each $\mathbf{0}$ value in the SAP-array denotes the start of a new *SAP-interval* in the BWT, within which all characters share an identical associated suffix.

The BWTs of $S$ and $S'$ can only differ within SAP-intervals that contains more than one distinct symbol. Within such an SAP-interval, the ordering of the characters is entirely determined by the ordering of the reads they are contained in, so best compression is achieved if we permute the characters so they are grouped into as few runs as possible. In doing this we are implicitly permuting the sequences in the collection. As illustrated by Figure 1, if the collection is sorted such that the reverses of the sequences are in lexicographic order (we call this *reverse lexicographic order*, or *RLO* for short), then the symbols in an SAP-interval of length $l$ must group into at most $\phi$ runs, where $\phi \leq \sigma$ is the number of distinct characters encountered within the interval. This compares with an upper bound up to $l$ runs if no sorting is applied. We would therefore expect the RLO-sorting of a collection to compress better than the original and our experiments in the next section show this to be the case in practice.

However, we wish our approach to scale to many millions of reads and so we would prefer to avoid sorting the collection as a preprocessing step. If we know the SAP-array of a BWT, it is a simple single-pass procedure to read both the BWT and its SAP-array from disk in tandem, identify SAP-intervals, sort the characters within them and output a modified BWT. We note there is scope to experiment with different heuristics - *e.g.* in Figure 1, placing the $\mathtt{T}$ prior to the two $\mathtt{G}$s in the $\mathtt{ACCT}$-interval would eliminate another run by extending the run of $\mathtt{T}$s begun in the preceding $\mathtt{ACCAT}$-interval.

It remains to compute the SAP-array. In order to do this, we show that the method of BWT construction introduced by three of the present authors in Bauer *et al.* (2011) can be modified to compute the SAP-array alongside the BWT with minimal additional overhead. We do not attempt to describe this previous work in full detail, but the BWT construction algorithm proceeds in $k$ stages, $k$ being the length of the reads in the collection.[3] At stage $j$, the $j$-suffixes ($0 \leq j \leq k$) of the reads (that is, the suffixes of length $j$, the 0-*suffix* being defined as the suffix that contains only the end marker \$) are processed in lexicographic order and the characters that precede them are merged into a 'partial BWT'.

The partial BWT at step $j$ can be thought of as the concatenation of $\sigma + 1$ segments $B_j(0), B_j(1), \ldots, B_j(\sigma)$, where each $B_j(h)$ corresponds to the symbols in the partial BWT segments that precede all suffixes of $S$ that are of length smaller or equal to

$j$ starting with $c_0 = \$^4$, for $h = 0$, and with $c_h \in \Sigma$, for $h = 1, \ldots, \sigma$.

At each step $j$, with $0 < j \leq k$ the main idea is to compute the positions of all characters that precede the $j$-suffixes in the old partial BWT in order to obtain an updated partial BWT that also contains the symbols associated with the suffixes of length $j$. The two variants $\mathtt{BCR}$ and $\mathtt{BCRext}$ of the Bauer *et al.* (2011) algorithm contrive in slightly different ways (with different trade-offs between RAM use and I/O) to arrange matters so that the symbols we have to insert into the partial BWT are processed by lexicographic order of their associated $j$-suffixes, which allows each segment of the partial BWT to be held on disk and accessed sequentially. Crucially, this also means that the characters in an SAP-interval will be processed consecutively. When we write a BWT character, we can also write its SAP status at the same time, since it is not going to change in subsequent iterations.

We consider two suffixes $u$ and $v$ of length $j$ of two different reads $s_r$ and $s_t$ satisfying $1 \leq r, t \leq m$ and $r \neq t$. Both $s_r$ and $s_t$ belong to $S$ and assume that $u$ is lexicographically less than $v$. Then, they are identical (up to the end-markers) and so the SAP status of $v$ must be set to 1 if and only if the following conditions hold: the symbols associated with suffixes $u'$ and $v'$ of length $(j-1)$ of $s_r$ and $s_t$ are equal which guarantees that $u$ and $v$ begin with the same symbol, and they are stored in the same BWT segment, implying that $u'$ and $v'$ begin with the same symbol. The SAP status of all suffixes between $u'$ and $v'$ must be set to 1 which ensures that all suffixes between $u'$ and $v'$ in iteration $j-1$ coincide and have the length $j-1$. Note that the symbols in the BWT segment associated with suffixes between $u'$ and $v'$ could be different from the symbol preceding $u'$ (and $v'$).

Actually, we compute the SAP status of the suffixes $u$ and $v$ at the step $j-1$ when we insert the corresponding suffixes $u'$ and $v'$. In particular, when we copy the values from the old to the new partial BWT and insert the new $m$ values, we can read the SAP values and verify the condictions stated above at the same time. It means that at the end of the iteration $j-1$, we obtain the partial BWT and the SAP values of the next suffixes to insert and, at iteration $j$, we can directly set the SAP status of the $j$-suffixes and compute the SAP status for the suffixes of the next iteration.

We can compute the SAP-interval by induction, *i.e.* we can compute the SAP status of the $j$-suffixes at iteration $j-1$ by using the symbols associated with the $(j-1)$-suffixes and their SAP values. At iteration $j-1$, we assume that we know the SAP values of the $(j-1)$-suffixes and we have to compute the SAP values of the $j$-suffixes need for the next iteration. For simplicity, we focus on the computation of the SAP values of a fixed BWT segment, *i.e.* we only consider the insertion of the $(j-1)$-suffixes starting with the same symbol. In our implementation, we use a counter $A$ for each symbol of the alphabet and a generic counter $Z$.

The element $A[h]$, for each $h = 1, \ldots, \sigma$ and $c_h \in \Sigma$ (we can ignore the end-marker because the reads have the same length and hence it does not appear in the partial BWT), contains the number of SAP intervals between the first position and the position of the last inserted symbol associated with a read $s_q$ (for some $1 \leq q \leq m$) equal to $c_h$ in the considered BWT segment. The counter $Z$ contains

---

the number of the SAP intervals between the first position and the position where we have to insert $c_p$. The symbol $c_p$ is associated with the new suffix of length $j - 1$ of read $s_t$, with $1 \le t \le m$, that we want to insert. If the value $A[p]$ is equal to $Z$, then the SAP status of the $j$-suffix of $s_t$ (obtained by concatenating $c_p(j-1)$-suffix of $s_t$) must be set to 1, otherwise it is set to 0. We observe that if $A[p] = Z$ holds true, then this implies that $j$-suffixes of $s_r$ and $s_t$ are in the same SAP interval.

| $B(0)$ | $SAP(0)$ | suffixes | $B_{SAP}(0)$ |
|---|---|---|---|
| $T$ | **0** | $\$_1$ | $T$ |
| $T$ | **1** | $\$_2$ | $T$ |
| $T$ | **1** | $\$_3$ | $T$ |
| $T$ | **1** | $\$_4$ | $T$ |
| $B(1)$ | $SAP(1)$ | suffixes | $B_{SAP}(1)$ |
| $T$ | 0 | $ACCACT\$_2$ | $T$ |
| **G** | **0** | **ACCT$\$_1$** | **G** |
| **T** | **1** | **ACCT$\$_2$** | **G** |
| **G** | **1** | **ACCT$\$_4$** | **T** |
| $C$ | 0 | $ACT\$_3$ | $C$ |
| **T** | **0** | **AGACCT$\$_1$** | **G** |
| **G** | **1** | **AGACCT$\$_4$** | **T** |
| $G$ | 0 | $ATACCT\$_2$ | $G$ |
| $B(2)$ | $SAP(2)$ | suffixes | $B_{SAP}(2)$ |
| $C$ | 0 | $CACT\$_3$ | $C$ |
| $A$ | 0 | $CCACT\$_3$ | $A$ |
| **A** | **1** | **CCT$\$_1$** | **A** |
| **A** | **1** | **CCT$\$_2$** | **A** |
| **A** | **1** | **CCT$\$_4$** | **A** |
| **C** | **0** | **CT$\$_1$** | **A** |
| **C** | **1** | **CT$\$_2$** | **C** |
| **A** | **1** | **CT$\$_3$** | **C** |
| **C** | **1** | **CT$\$_4$** | **C** |
| $B(3)$ | $SAP(3)$ | suffixes | $B_{SAP}(3)$ |
| **A** | **0** | **GACCT$\$_1$** | **A** |
| **A** | **1** | **GACCT$\$_4$** | **A** |
| $\$_4$ | 0 | $GAGACCT\$_4$ | $\$_4$ |
| $\$_2$ | 0 | $GATACCT\$_2$ | $\$_2$ |
| $B(4)$ | $SAP(4)$ | suffixes | $B_{SAP}(4)$ |
| **C** | **0** | **T$\$_1$** | **C** |
| **C** | **1** | **T$\$_2$** | **C** |
| **C** | **1** | **T$\$_3$** | **C** |
| **C** | **1** | **T$\$_4$** | **C** |
| $\$_3$ | 0 | $TACCACT\$_3$ | $\$_3$ |
| $A$ | 0 | $TACCT\$_2$ | $A$ |
| $\$_1$ | 0 | $TAGACCT\$_1$ | $\$_1$ |

$\Rightarrow$

**Fig. 1.** Columnwise from left, we have the BWT of the collection $S = \{TAGACCT, GATACCT, TACCACT, GAGACCT\}$, the SAP bit and the suffix associated with each symbol. The right-hand side shows the BWT of the collection $\{TACCACT, TAGACCT, GAGACCT, GATACCT\}$ obtained by sorting the elements of $S$ into reverse lexicographical order (RLO). This permutes the symbols within SAP-intervals so as to minimise the number of runs.

## 3 RESULTS

Reads simulated from the *E.coli* genome (K12 strain) allowed us to assess separately the effects of coverage, read length and sequencing error on the level of compression achieved. First, a $60\times$ coverage of error-free 100 base reads was subsampled into datasets as small as $10\times$. Figure 2 shows a summary plot of the compression ratios at various coverage levels for compression both on the original reads and the BWT transform.

We found the PPMd mode (`-m0=PPMd`) of `7-Zip` to be a good choice of second-stage compressor for the BWT strings (referred to as *PPMd (default)* in the following). RLO-sorting the datasets led to a BWT that was slightly more compressible than the SAP-permuted BWT, but the difference was small (0.36bpb versus 0.38bpb at $60\times$, both over 30% less than the 0.55bpb taken up by the compressed BWT of the unsorted reads).

In contrast, when `gzip`[5], `bzip2` and default PPMd were applied to the original reads, each gave a compression level that was consistent across all levels of coverage and none was able to compress below 2 bits per base. However, a sweep of the PPMd parameter space yielded a combination `-mo=16 -mmem=2048m` that attained 0.50bpb on the $60\times$ dataset (in the following we will refer to this parameter setting as *PPMd (large)*). This is because the *E.coli* genome is small enough to permit several-fold redundancy of the genome to be captured in the 2Gbytes of working space that this combination specifies. For a much larger genome such as human, this advantage disappears. Figure 3 summarizes results from a set of 192 million human reads[6] previously analyzed by Yanovsky (2011): PPMd(large) compresses the BWT of the reads less well than PPMd(default), as well as being several times slower.

We also investigated the effects of sequencing errors on the compression ratios by simulating $40\times$ data sets of 100bp reads with different rates of uniformly distributed substitution error, finding that an error rate of 1.2% approximately doubled the size of the compressed BWT (0.90bpb, compared with 0.47bpb for error-free data at the same coverage).

We were interested in the behaviour of BWT-based compression techniques as a function of the read length. To this end, we fixed a coverage of $40\times$ and simulated error-free *E.coli* reads of varying lengths. As the read length increased from 50bp to 800bp, the size of the compressed BWTs shrank from 0.54bpb to 0.32bpb. This is not surprising since longer reads allow repetitive sequences to be grouped together, which could otherwise potentially be disrupted by suffixes of homologous sequences.

Finally, we assessed the performance of the compressors on a typical human genome resequencing experiment[7], containing 135.3Gbp of 100-base reads, or about $45\times$ coverage of the human genome. In addition to this set of reads, we created a second dataset by trimming the reads based on their associated quality scores according to the scheme described in `bwa` (Li and Durbin (2009)). Setting a quality score of 15 as the threshold removes 1.3% of the input bases. We again constructed the corresponding data sets in RLO and SAP order. Table 4 shows the improvement in terms of compression after the trimming. Eliminating 1.3% of the bases improves the compression ratio by about 4.5%, or compressing the entire 133.6Gbp down to 7.7Gbytes.

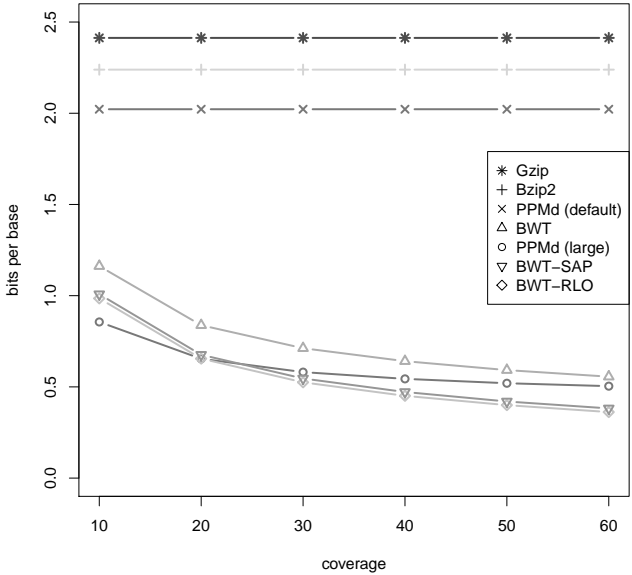

**Fig. 2.** We simulated $60\times$ coverage of error-free 100 base reads from the *E.coli* genome and subsampled this into datasets as small as $10\times$. We compared different compression schemes, both on the raw input sequences and the BWT. RLO denotes a reverse lexicographical ordering of the reads, SAP is the data set where all the reads are ordered according to the same-as-previous array. The $x$-axis gives the coverage level whereas the $y$-axis shows the number of bits used per input symbol. Gzip, Bzip2, PPMd (default) and PPMd (large) show compression achieved on the raw sequence data. BWT, BWT-SAP and BWT-RLO give compression results on the BWT using PPMd (default) as second-stage compressor.

| Method | | Time | | Compression |
|---|---|---|---|---|
| Stage 1 | Stage 2 | Stage 1 | Stage 2 | |
| Reads | Bzip2 | - | 905 | 2.25 |
| | PPMd (default) | | 324 | 2.04 |
| | PPMd (large) | | 5155 | 2.00 |
| | -mx9 | | 17974 | 1.98 |
| BWT | Bzip2 | 3520 | 818 | 2.09 |
| | PPMd (default) | | 353 | 1.93 |
| | PPMd (large) | | 4953 | 2.05 |
| | -mx9 | | 16709 | 2.09 |
| BWT-SAP | Bzip2 | 3520 | 601 | 1.40 |
| | PPMd (default) | | 347 | 1.21 |
| | PPMd (large) | | 3116 | 1.28 |
| | -mx9 | | 11204 | 1.34 |

**Fig. 3.** Different combinations of first-stage (BWT, SAP-permuted BWT) and second-stage (bzip2 with default parameters, PPMd mode of `7-Zip` with default parameters, PPMd mode of `7-Zip` with `-mo=16 -mmem=2048m`, deflate mode of `7-Zip` with `-mx9`) compression compared on 192 million human reads previous analyzed by Yanovsky (2011). Time is in CPU seconds, as measured on a single core of an Intel Xeon X5450 (Quad-core) 3GHz processor. Compression is in bits per base.

| | Input size | BWT | BWT-RLO | BWT-SAP |
|---|---|---|---|---|
| **untrimmed** | 135.3Gbp | 0.746 | 0.528 | 0.484 |
| **trimmed** | 133.6Gbp | 0.721 | 0.504 | 0.462 |

**Fig. 4.** Different combinations of first-stage (BWT, SAP- and RLO-permuted BWT) and PPMd as the second-stage compression method compared on a $45\times$ human data set. Values state the compression achieved in bits per input base. We trimmed the reads by following the strategy described for `bwa` which removed 1.3% of the bases. This leads to an improvement in terms of compression ratio of 4.5%.

## 4 DISCUSSION

Analysing the SRX001540 dataset allowed us to compare our results with ReCoil, but the reads are noisier and shorter than more recent datasets and, at under $3\times$, the oversampling of the genome is too low to manifest the kind of coverage-induced compression that Figure 2 illustrates. Nevertheless, we were able to compress the data to 1.21bpb in just over an hour, compared to 1.34bpb achieved by ReCoil in 14 hours (albeit on a slower processor). On the more recent ERA015743 data we obtained 0.48bpb, allowing the 135.3Gbp of data to fit in 8.2Gbytes of space.

In both cases, reordering the strings in the collection - implicitly or explicitly - is necessary for best compression. One situation where this might be problematic would be if the reads are paired - typically, sequenced from different ends of a library of DNA templates. An obvious way to capture this relationship would be to store the two reads in a pair at adjacent positions in $S$ - i.e. the first two reads in the collection are reads 1 and 2 of the first read-pair, and so forth. If we wanted to retain this information after a change in ordering, we would incur the overhead of storing an additional pointer to associate each read with its mate.

In Bauer *et al.* (2012), we give efficient algorithms for inverting the BWT and recovering the original reads. However, when augmented with relatively small additional data structures, the compressed BWT forms the core of the FM-index (Ferragina and Manzini (2000, 2005); Ferragina *et al.* (2007)) that allows `count` ("how many times does a $k$-mer occur in the dataset?") and `locate` ("at what positions in the collection does the $k$-mer occur?") queries. Several variants of this algorithmic scheme exist which make different tradeoffs between the compression achieved and the time needed for the `count` and `locate` operations. For instance, Ferragina *et al.* (2007) describes a variant of the FM-index that indexes a string $T$ of length $n$ within $nH_k(T)+o(n)$ bits of storage, where $H_k(T)$ is the $k$-th order empirical entropy of $T$. This index counts the occurrences in $T$ of a string of length $p$ in $O(p)$ time and it locates each occurrence of the query string in $O(\log^{1+\epsilon} n)$ time, for any constant $0 < \epsilon < 1$.

Most sequencing technologies associate a quality score with each sequenced base and Kozanitis *et al.* (2010) found that significant lossless compression of these scores can be hard to achieve, but also showed that a lossy reduction in resolution of the scoring scheme could obtain better compression while having limited impact on the accuracy of variant calls derived from the sequences. Future work we are pursuing would complement such approaches by using queries to a BWT-based index for *de novo* identification of bases that are not likely to be important in downstream variant calls, whose quality scores can then be discarded entirely.

More broadly, many computationally intensive tasks in sequence analysis might be facilitated by having a set of reads in indexed form. For example, Simpson and Durbin (2010, 2011) show how the overlap of a set of reads can be computed from its FM-index and apply this insight to build a practical tool for *de novo* assembly. By making it feasible to index sets of reads on the scale necessary for whole human genome sequencing, our work raises the intriguing possibility of a new generation of algorithms that operate directly on sets of reads held in compressed index form.

## ACKNOWLEDGEMENT

The authors would like to thank Dirk Evers for his support throughout this project.

*Funding*: M.J.B. and A.J.C. are employees of Illumina Inc., a public company that develops and markets systems for genetic analysis, and receive shares as part of their compensation. Part of T.J.'s contribution was made while on a paid internship at Illumina's offices in Cambridge, UK.

## REFERENCES

Adjeroh, D., Bell, T., and Mukherjee, A. (2008). *The Burrows-Wheeler Transform: Data Compression, Suffix Arrays, and Pattern Matching*. Springer Publishing Company, Incorporated, 1st edition.

Bauer, M. J., Cox, A. J., and Rosone, G. (2011). Lightweight BWT construction for very large string collections. In *CPM 2011*, volume 6661 of *LNCS*, pages 219–231. Springer.

Bauer, M. J., Cox, A. J., and Rosone, G. (2012). Lightweight algorithms for constructing and inverting the BWT of string collections. *Theoretical Computer Science*. Available online 10 February 2012.

Burrows, M. and Wheeler, D. J. (1994). A block sorting data compression algorithm. Technical report, DIGITAL System Research Center.

Chen, X., Li, M., Ma, B., and Tromp, J. (2002). DNACompress: fast and effective DNA sequence compression. *Bioinformatics*, **18**(12), 1696–1698.

Deorowicz, S. and Grabowski, S. (2011). Compression of genomic sequences in FASTQ format. *Bioinformatics*, **27**(6), 860–862.

Dewey, F. E., Chen, R., Cordero, S. P., Ormond, K. E., Caleshu, C., Karczewski, K. J., Whirl-Carrillo, M., Wheeler, M. T., Dudley, J. T., Byrnes, J. K., Cornejo, O. E., Knowles, J. W., Woon, M., Sangkuhl, K., Gong, L., Thorn, C. F., Hebert, J. M., Capriotti, E., David, S. P., Pavlovic, A., West, A., Thakuria, J. V., Ball, M. P., Zaranek, A. W., Rehm, H. L., Church, G. M., West, J. S., Bustamante, C. D., Snyder, M., Altman, R. B., Klein, T. E., Butte, A. J., and Ashley, E. A. (2011). Phased whole-genome genetic risk in a family quartet using a major allele reference sequence. *PLoS Genet*, **7**(9), e1002280.

Ferragina, P. and Manzini, G. (2000). Opportunistic data structures with applications. In *Proceedings of the 41st Annual Symposium on Foundations of Computer Science*, pages 390–398, Washington, DC, USA. IEEE Computer Society.

Ferragina, P. and Manzini, G. (2005). Indexing compressed text. *J. ACM*, **52**, 552–581.

Ferragina, P., Manzini, G., Mäkinen, V., and Navarro, G. (2007). Compressed representations of sequences and full-text indexes. *ACM Trans. Algorithms*, **3**.

Fritz, M. H., Leinonen, R., Cochrane, G., and Birney, E. (2011). Efficient storage of high throughput DNA sequencing data using reference-based compression. *Genome Research*, **21**(5), 734–740.

Giancarlo, R., Scaturro, D., and Utro, F. (2009). Textual data compression in computational biology: a synopsis. *Bioinformatics*, **25**(13), 1575–1586.

Grumbach, S. and Tahi, F. (1994). A new challenge for compression algorithms: Genetic sequences. In *Information Processing and Management*, volume 30, pages 875–886.

Kozanitis, C., Saunders, C., Kruglyak, S., Bafna, V., and Varghese, G. (2010). Compressing genomic sequence fragments using SlimGene. In *RECOMB*, volume 6044 of *LNCS*, pages 310–324. Springer.

Li, H. and Durbin, R. (2009). Fast and accurate short read alignment with Burrows-Wheeler transform. *Bioinformatics*, **25**(14), 1754–1760.

Mantaci, S., Restivo, A., Rosone, G., and Sciortino, M. (2005). An Extension of the Burrows Wheeler Transform and Applications to Sequence Comparison and Data Compression. In *CPM 2005*, volume 3537 of *LNCS*, pages 178—189.

Milosavljevic, A. and Jurka, J. (1993). Discovering simple DNA sequences by the algorithmic significance method. *Computer applications in the biosciences : CABIOS*, **9**(4), 407–411.

Rivals, E., Dauchet, M., Delahaye, J., and Delgrange, O. (1996). Compression and genetic sequence analysis. *Biochimie*, **78**(5), 315–22.

Simpson, J. T. and Durbin, R. (2010). Efficient construction of an assembly string graph using the FM-index. *Bioinformatics*, **26**(12), i367–i373.

Simpson, J. T. and Durbin, R. (2011). Efficient de novo assembly of large genomes using compressed data structures. *Genome Research*. Published in Advance December 7, 2011.

Tembe, W., Lowey, J., and Suh, E. (2010). G-SQZ: compact encoding of genomic sequence and quality data. *Bioinformatics*, **26**(17), 2192–2194.

Yanovsky, V. (2011). ReCoil - an algorithm for compression of extremely large datasets of DNA data. *Algorithms for Molecular Biology*, **6**(1), 23.